\documentstyle[12pt]{article}
\begin{document}
\begin{flushright}
 THES-TP 26/94\\
 DEM-NT  94-36\\
November 1994
\end{flushright}

\vskip 0.3in

\begin{center}{ \large\bf The  symmetry algebra of the  N-dimensional
anisotropic quantum harmonic oscillator with
 rational ratios of frequencies and the Nilsson model}

\end{center}

\bigskip\bigskip

\begin{center}
Dennis Bonatsos$^{\dagger,\ast}$\footnote{e:mail \begin{tabular}[t]{l}
bonat\@ nuclear.ect.unitn.it\\
bonat\@ cyclades.nrcps.ariadne-t.gr
\end{tabular}},
C.~Daskaloyannis$^{\dagger\dagger}$
\footnote{e:mail~~daskaloyanni@olymp.ccf.auth.gr},
 P.~Kolokotronis$^{\ast}$ and D. Lenis$^{\ast}$
\bigskip

$^\dagger$ European Centre for
 Theoretical Research in Nuclear Physics and Related
Areas (ECT$^*$)

Villa Tambosi, Strada delle Tabarelle 286, I-38050 Villazzano (Trento), Italy

$^\ast$ Institute of Nuclear Physics, N.C.S.R.
``Demokritos''

GR-15310 Aghia Paraskevi, Attiki, Greece

$^\dagger\dagger$ Department of Physics, Aristotle University of
Thessaloniki

GR-54006 Thessaloniki, Greece
\end{center}

\bigskip\bigskip\bigskip

\centerline{\bf Abstract}

The symmetry algebra of the N-dimensional anisotropic quantum harmonic
oscillator with rational ratios of frequencies is constructed by a method
of general applicability to quantum superintegrable systems. The special
case of the 3-dim oscillator is studied in more detail, because of its
relevance in the description of superdeformed nuclei and nuclear and atomic
clusters. In this case the symmetry algebra turns out to be a nonlinear
extension of the u(3) algebra. A generalized angular momentum operator
useful for labeling the degenerate states is constructed, clarifying the
connection of the present formalism to the Nilsson model in nuclear physics.

\bigskip\bigskip


\newpage
 {\bf 1. Introduction }

Quantum algebras \cite{Dri,Jim} (also called quantum groups) are nonlinear
generalizations of the usual Lie algebras, to which they reduce in the
limiting case in which the deformation parameters are set equal to unity.
{}From the mathematical point of view they have the structure of Hopf
algebras \cite{Abe}. The interest for applications of quantum algebras in
physics was triggered in 1989 by the introduction of the q-deformed
harmonic oscillator \cite{Bie,Mac,Sun}, although such mathematical structures
have already been in existence \cite{Ari,Kur}. The q-deformed harmonic
oscillator has been introduced as a tool for providing a boson realization
for the quantum algebra su$_q$(2). Since then several generalized deformed
oscillators (see \cite{Das,Arik,Brz,PLB307,Mel} and
references therein), as well as generalized deformed versions of the su(2)
algebra \cite{Kol,DQ,LG,SL,Pan} have been introduced and used in several
applications to physical problems.

On the other hand, the 3-dimensional anisotropic quantum harmonic oscillator
is of interest in the description of many physical systems. Its single
particle level spectrum is known \cite{Mot,Rae} to characterize the
basic structure of superdeformed and hyperdeformed nuclei \cite{NT,JK}.
Furthermore, it has been recently connected \cite{RZ,ZRM} to the underlying
geometrical structure  in the Bloch--Brink $\alpha$-cluster model \cite{BB}.
It is also of interest for the interpretation of the observed shell
structure in atomic clusters \cite{Mar}, especially after the realization
that large deformations can occur in such systems \cite{Bul}.
The 2-dimensional anisotropic quantum harmonic oscillator is also of interest,
since, for example, its single particle level spectrum characterizes the
underlying symmetry of ``pancake'' nuclei \cite{Rae}.

The anisotropic harmonic oscillator in two \cite{JH,Dem,Con,Cis,GKM,CLP}
and three \cite{DZ,Mai,Ven,MV,RD,BCD,ND} dimensions has been the
subject of several investigations, both at the classical and at the quantum
mechanical level. The special cases with frequency ratios 1:2 \cite{Holt,BW}
and 1:3 \cite{FL} have also been considered. The anisotropic harmonic
oscillator is a well known example of a superintegrable system, both at
the classical and at the quantum mechanical level \cite{Hie,Eva}.
Although at the classical level it is clear that for
the N-dimensional anisotropic harmonic oscillator the su(N) or sp(2N,R)
algebras can be used for its description, the situation at the quantum
level is not equally clear \cite{JH}.

In section 2 of this letter the symmetry algebra of the N-dimensional
anisotropic quantum harmonic oscillator with rational ratios of frequencies
is studied by a method of general applicability to superintegrable systems
\cite{BDK}, while in section 3 the case of the 3-dimensional oscillator
is considered in more detail, because of its importance for the description
of nuclei and possibly of atomic clusters. In the  case of the 3-dim oscillator
the symmetry
algebra turns out to be a nonlinear generalization of u(3). In addition
a generalized angular momentum operator is constructed, suitable for labelling
the degenerate states in a way similar to that used in the framework of the
Nilsson model \cite{Nil}.

{\bf 2. General case: the N-dimensional  oscillator}

Let us consider the system described by the Hamiltonian:
\begin{equation}
H=\frac{1}{2}
\sum\limits_{k=1}^N \left(
{p_k}^2 + \frac{x_k^2}{m_k^2}
 \right),
\label{eq:Hamiltonian}
\end{equation}
where $m_i$ are natural numbers mutually prime ones, i.e.
their great common divisor is $\gcd (m_i,m_j)=1$ for $i\ne j$ and
$i,j=1,\ldots,N$.

We define the creation and annihilation operators \cite{JH}:
\begin{equation}
\begin{array}{ll} a_k^\dagger=& \frac{1}{\sqrt{2}}
\left(\frac{x_k}{m_k} -i {p_k}  \right), \\
a_k=&\frac{1}{\sqrt{2}}
\left(\frac{x_k}{m_k} +i {p_k}  \right), \\
U_k=&
\frac{1}{2}
\left(
{p_k}^2 + \frac{x_k^2}{m_k^2}
\right)= \frac{1}{2} \left\{ a_k, a_k^\dagger \right\},
\quad H= \sum\limits_{k=1}^N U_k.
\end{array}
\label{eq:operators}
\end{equation}

These operators satisfy the relations (the indices $k$ have been omitted):
\begin{equation}
\begin{array}{ll}
a^\dagger\, U &= \left( U- \frac{1}{m} \right) a^\dagger \quad {\rm or} \quad
[U, \left(a^\dagger\right)^m ]= \left( a^\dagger \right)^m, \\
a\, U &= \left( U+ \frac{1}{m} \right)a \quad {\rm or}
 \quad [U, \left(a\right)^m ]= -\left(a\right)^m, \\
a^\dagger a &= U-\frac{1}{2m}, \qquad
a a^\dagger =U + \frac{1}{2m} , \\
\left[ a, a^{\dagger} \right] &= \frac{1}{m} , \\
\left(a^\dagger\right)^m \left(a\right)^m &=F(m,U), \\
\left(a\right)^m \left(a^\dagger\right)^m &=F(m,U+1), \\
\end{array}
\label{eq:basic}
\end{equation}
where the function $F(m,x)$ is defined by:
\begin{equation}
F(m,x)=\prod\limits_{p=1}^m{\left( x - \frac{2p-1}{2m}\right)\,}.\label{eq:F}
\end{equation}

Using the above relations we can define the enveloping
 algebra ${\cal C}$, defined by the polynomial combinations of the
generators $\Big\{ 1, H, {\cal A}_k, {\cal A}_k^\dagger, U_k
\Big\}$ and $k=1,\ldots, N-1$,  where:
\begin{equation}
{\cal A}_k^\dagger= \left( a_k^\dagger\right)^{m_k} \left( a_N \right)^{m_N},
\qquad
{\cal A}_k= \left( a_k\right)^{m_k} \left( a_N ^\dagger\right)^{m_N}.
\end{equation}
These operators correspond to a multidimensional generalization of eq.
(\ref{eq:operators}):
\begin{equation}
\big[ H, {\cal A}_k \big]=0,\quad
\big[ H, {\cal A}_k^\dagger \big]=0,\quad
\big[H, {U}_k  \big]=0,\quad
k=1,\ldots, N-1.
\label{eq:super}
\end{equation}
The following relations are satisfied for $k\ne \ell$
and $k,\ell =1,\ldots,N-1$:
\begin{equation}
\left[{U}_k, {\cal A}_\ell \right]=
\left[{U}_k, {\cal A}_\ell^\dagger \right]=
\left[{\cal A}_k, {\cal A}_\ell \right]=
\left[{\cal A}_k^\dagger, {\cal A}_\ell^\dagger \right]=0,
\label{eq:different}
\end{equation}
while
\begin{equation}
\begin{array}{l}
{U}_k{\cal A}_k^\dagger=  {\cal A}_k^\dagger \left( {U}_k +1 \right), \\
{U}_k {\cal A}_k=  {\cal A}_k\left( {U}_k -1 \right), \\
{\cal A}_k^\dagger{\cal A}_k=F\Big( m_k, {U}_k \Big) F\Big(m_N,  H
- \sum\limits_{\ell=1}^{N-1}{U}_\ell\, +1\Big),  \\
{\cal A}_k{\cal A}_k^\dagger=F\Big( m_k,   {U}_k +1 \Big) F\Big(m_N,  H
- \sum\limits_{\ell=1}^{N-1}{U}_\ell\Big). \\
\end{array}
\label{eq:kk}
\end{equation}
One  additional relation for $k\ne \ell$ can be derived:
\begin{equation}
F\Big(m_N,  H  - \sum\limits_{\ell=1}^{N-1}{U}_\ell\Big) {\cal A}_k^\dagger
{\cal A}_\ell =
F\Big(m_N,  H  - \sum\limits_{\ell=1}^{N-1}{U}_\ell\, +1\Big)
{\cal A}_\ell {\cal A}_k^\dagger.
\label{eq:kl}
\end{equation}

The algebra defined by the above relations accepts a Fock space
 representation. The elements of the basis
$\left\vert E, p_1,\ldots,p_{N-1} \right>$
are characterized by the
eigenvalues of the $N$
commuting elements  of the algebra $H$ and ${U}_k$ with
$k=1,\ldots,N-1$. The elements ${\cal A}_k$ and ${\cal A}_k^\dagger$
are the corresponding ladder operators of the algebra. The following
relations hold:

\begin{equation}
\begin{array}{l}
H \left\vert E, p_1,\ldots,p_{N-1} \right>=
E\left\vert E, p_1,\ldots,p_{N-1} \right>, \\
{U}_k \left\vert E, p_1,\ldots,p_{N-1} \right>=
p_k\left\vert E, p_1,\ldots,p_{N-1} \right>, \\
{\cal A}_k \left\vert E, p_1,\ldots,p_k,\ldots,p_{N-1} \right>=\\
\sqrt{F\Big( m_k,   { p}_k\Big) F\Big(m_N,  E
- \sum\limits_{\ell=1}^{N-1}{p}_\ell+1\Big) }
\left\vert E, p_1,\ldots,p_k-1,\ldots,p_{N-1} \right>, \\
{\cal A}_k^\dagger \left\vert E, p_1,\ldots,p_k,\ldots,p_{N-1} \right>=\\
\sqrt{F\Big( m_k,   { p}_k+1 \Big) F\Big(m_N,  E
- \sum\limits_{\ell=1}^{N-1}{p}_\ell\Big)}
\left\vert E, p_1,\ldots,p_k+1,\ldots,p_{N-1} \right>. \\
\end{array}
\label{eq:repr}
\end{equation}

Let $p_k^{\mbox{min}}$ be the minimum value of $p_k$ such that
\begin{equation}
{\cal A}_k \left\vert E, p_1,\ldots,p_k^{\mbox{min}},\ldots,p_{N-1}
 \right>=0.
\label{eq:maxweight}
\end{equation}
{}From eq. (\ref{eq:repr}) we find that we must have:
\begin{equation}
F\Big( m_k,{ p}_k^{\mbox{min}} \Big)=0. \label{eq:1}
\end{equation}
Then
 $p_k$ is one of the roots of the function $F$ defined by
eq. (\ref{eq:F}).
The general form of the roots is:
\begin{equation}
p_k^{\mbox{min}}=\frac{2q_k -1 }{2 m_k }, \qquad q_k= 1,\ldots,m_k.
\label{eq:roots}
\end{equation}
Each root is characterized by a number $q_k$.  The numbers $q_k$
 also characterize the representations of the algebra, as we shall
see.

The elements of the Fock space can be generated by successive
applications of the ladder operators ${\cal A}_k^\dagger$ on the
minimum weight element
\begin{equation}
\left\vert E, p_1^{\mbox{min}},\ldots, p_{N-1}^{\mbox{min}}\right>=\left\vert
\begin{array}{c} E, \left[0\right]\\ \left[ q \right] \end{array}
\right>=\left\vert \begin{array}{c} E, 0, \ldots , 0 \\ q_1,
\ldots,q_k,\ldots,q_{N} \end{array} \right>.
\label{eq:min}
\end{equation}
The elements of the basis of the Fock space are given by:
\begin{equation}
\left\vert E, [p_k^{\mbox{min}}+n_k] \right>=
\left\vert \begin{array}{c} E, \left[n\right]\\ \left[ q \right] \end{array}
\right>
= \frac{1}{\sqrt{C^{[n]}_{[q]}}} \left(
\prod\limits_{k=1}^{N-1}\left({\cal A}_k^\dagger\right)^{n_k} \right)
\left\vert\begin{array}{c} E, \left[0\right]\\  \left[ q \right]  \end{array}
\right>,
\label{eq:base}
\end{equation}
where $[n]= (n_1, n_2, \cdots, n_{N-1})$ and $[q]= (q_1,q_2, \ldots, q_{N})$,
while $C^{[n]}_{[q]}$ are normalization coefficients.

The generators of the algebra acting on the base  of the Fock space give:
\begin{equation}
\begin{array}{l}
H \left\vert\begin{array}{c}E, \left[n\right]\\ \left[ q \right] \end{array}
\right>=E \left\vert \begin{array}{c} E, \left[n\right]\\ \left[ q \right]
\end{array} \right>, \\
{U}_k \left\vert \begin{array}{c} E, \left[n\right]\\ \left[ q \right]
\end{array} \right> = \left(n_k + p_k^{\mbox{min}}\right) \left\vert
\begin{array}{c} E, \left[n\right]\\ \left[ q \right] \end{array} \right>,
\quad k=1,\ldots,N-1, \\ \end{array}
\label{eq:diag}
\end{equation}
\begin{equation}
\begin{array}{l}
{\cal A}^\dagger_k
\left\vert
\begin{array}{c}
E, \left[n\right]\\
\left[ q \right]
\end{array}
\right>
=\\
=
\sqrt{
F\Big( m_k,  n_k+p_k^{\mbox{min}} +1 \Big)
F\Big(m_N,  E
- \sum\limits_{\ell=1}^{N-1}\left(n_\ell+p_\ell^{\mbox{min}}\right)\Big)
}
\cdot\\
\quad \cdot
\left\vert
\begin{array}{c}
E, n_1,\ldots, n_k+1,\ldots,n_{N-1}\\
 q_1,\ldots,q_k,\ldots,
q_N
\end{array}
\right>,
\end{array}
\label{eq:Ak+}
\end{equation}
\begin{equation}
\begin{array}{l}
{\cal A}_k
\left\vert
\begin{array}{c}
E, \left[n\right]\\
\left[ q \right]
\end{array}
\right>
=\\
=
\sqrt{
F\Big( m_k,   n_k+p_k^{\mbox{min}} \Big)
F\Big(m_N,  E
- \sum\limits_{\ell=1}^{N-1}\left(n_\ell+p_\ell^{\mbox{min}}\right)+1\Big)
}
\cdot\\
\quad \cdot
\left\vert
\begin{array}{c}
E, n_1,\ldots, n_k-1,\ldots,n_{N-1}\\
 q_1,\ldots,q_k,\ldots,
q_N
\end{array}
\right>.
\end{array}
\label{eq:Ak}
\end{equation}
The existence of a finite dimensional representation implies
that after $\Sigma$ successive applications of the ladder
operators ${\cal A}^\dagger$ on the minimum weight element one gets zero,
so that the following condition is satisfied:
$$ F(m_N,E-\Sigma - \sum\limits_{\ell=1}^{N-1}p_\ell^{\mbox{min}})=0. $$
Therefore:
$$E -\Sigma - \sum\limits_{\ell=1}^{N-1}p_\ell^{\mbox{min}}=p_N^{\mbox{min}},$$
where $p_N^{\mbox{min}}$ is the root of equation
$F(m_N,p_N^{\mbox{min}})=0$. Then
\begin{equation}
E= \Sigma + \sum\limits_{k=1}^{N} \frac{2q_k-1}{2 m_k}\label{eq:energy}.
\end{equation}
In the case of finite dimensional representations  only the energies given
by eq.  (\ref{eq:energy}) are permitted and the elements of the
Fock space can be described by using $\Sigma$ instead of  $E$.
The action of the generators on the Fock space is described by
the following relations:
\begin{equation}
\begin{array}{l}
H\left\vert \begin{array}{c} \Sigma, \left[n\right]\\ \left[ q \right]
\end{array} \right>=  \left( \Sigma + \sum\limits_{k=1}^{N}
\frac{2q_k-1}{2 m_k} \right) \left\vert \begin{array}{c} \Sigma,
\left[n\right]\\ \left[ q \right] \end{array} \right>, \\
{U}_k \left\vert \begin{array}{c} \Sigma, \left[n\right]\\
\left[ q \right]
\end{array}
\right>
=
\left(n_k + \frac{2q_k-1}{2 m_k}\right)
\left\vert
\begin{array}{c}
\Sigma, \left[n\right]\\
\left[ q \right]
\end{array}
\right>, \quad k=1,\ldots,N-1, \\
\end{array}
\label{eq:diag1}
\end{equation}
\begin{equation}
\begin{array}{l}
{\cal A}^\dagger_k
\left\vert
\begin{array}{c}
\Sigma, \left[n\right]\\
\left[ q \right]
\end{array}
\right>
=\\
=
\sqrt{
F\Big( m_k,   n_k
+\frac{2q_k-1}{2 m_k} +1 \Big)
F\Big(m_N,  \Sigma
- \sum\limits_{\ell=1}^{N-1}n_\ell\,
+\frac{2q_N-1}{2 m_N}\Big)
}
\cdot\\
\quad \cdot
\left\vert
\begin{array}{c}
\Sigma, n_1,\ldots, n_k+1,\ldots,n_{N-1}\\
 q_1,\ldots,q_k,\ldots,
q_N
\end{array}
\right>,
\end{array}
\label{eq:Ak1+}
\end{equation}
\begin{equation}
\begin{array}{l}
{\cal A}_k
\left\vert
\begin{array}{c}
\Sigma, \left[n\right]\\
\left[ q \right]
\end{array}
\right>
=\\
=
\sqrt{
F\Big( m_k,   n_k+\frac{2q_k-1}{2 m_k} \Big)
F\Big(m_N,  \Sigma
- \sum\limits_{\ell=1}^{N-1}n_\ell\,+
\frac{2q_N-1}{2 m_N}+1\Big)
}
\cdot\\
\quad \cdot
\left\vert
\begin{array}{c}
\Sigma, n_1,\ldots, n_k-1,\ldots,n_{N-1}\\
 q_1,\ldots,q_k,\ldots,
q_N
\end{array}
\right>.
\end{array}
\label{eq:Ak1}
\end{equation}
The dimension of the representation is given by
$$ d= {\Sigma+N-1 \choose \Sigma }= {(\Sigma+1) (\Sigma+2)\cdots (\Sigma+N-1)
\over (N-1)!}.$$
It is clear that to each value of $\Sigma$ correspond $m_1 m_2 \dots m_N$
energy eigenvalues, each eigenvalue having degeneracy $d$.

Using eqs (\ref{eq:basic}) we can prove that
the algebra  generated by the generators
$a_\ell^\dagger,a_\ell, N_\ell=m_\ell U_\ell-1/2$ is an oscillator algebra
with structure function \cite{Das,PLB307}
 $$\Phi_\ell(x)= x/m_\ell, $$
i.e. with
$$ \Phi_l (N_l)= U_l-{1\over 2 m_l}.$$
This oscillator algebra is characterized by the commutation relations:
\begin{equation}
\left[ N_\ell, a_\ell^\dagger\right]=a^\dagger_\ell,
\quad
\left[ N_\ell, a_\ell\right]=- a_\ell,
\quad
a^\dagger_\ell a_\ell = \Phi_\ell\left( N_\ell \right),
\quad
a_\ell a_\ell^\dagger = \Phi_\ell\left( N_\ell +1 \right).
\label{eq:oscalgebra}
\end{equation}
There are in total $N$ different oscillators of this type, uncoupled
to each other. The  Fock space corresponding
 to these oscillators defines an infinite dimensional representation of
the algebra defined eqs (\ref{eq:super}-\ref{eq:kl}).
In order to see the connection of the present basis to the usual Cartesian
basis, one can use for the latter the symbol $[r]=(r_1,r_2,\ldots,r_N)$.
One then has
$$
\begin{array}{l}
a^\dagger_\ell
\left\vert [r] \right>=
\sqrt{\Phi(r_\ell+1)} \left\vert r_1,\ldots,r_\ell+1,\ldots,r_N \right>,
\\
a_\ell
\left\vert [r] \right>=
\sqrt{\Phi(r_\ell)} \left\vert r_1,\ldots,r_\ell-1,\ldots,r_N \right>,
\\
N_\ell
\left\vert [r] \right>=
r_\ell\left\vert [r] \right>.
\end{array}
$$
The connection between the above basis and the basis defined by eqs
(\ref{eq:diag1}-\ref{eq:Ak1}) is given by:
\begin{equation}
\begin{array}{c}
\left\vert [r] \right>=
\left\vert
\begin{array}{c}
\Sigma, \left[n\right]\\
\left[ q \right]
\end{array}
\right>, \\
\begin{array}{c}
r_\ell=n_\ell m_\ell + \mbox{mod}\left(r_\ell,m_\ell\right)\\
\ell = 1,\ldots, N
\end{array}
\longleftrightarrow
\begin{array}{l}
n_k=\left[ r_k/m_k \right]\\
k=1,\ldots,N-1\\
q_\ell=\mbox{mod }\left(r_\ell,m_\ell\right)+1\\
\Sigma = \sum\limits_{\ell=1}^N \left[ r_\ell/m_\ell\right]
\end{array}
\end{array}
\label{eq:corresp}
\end{equation}
where $[x]$ means the integer part of the number $x$.

Using the correspondence between the present basis and the usual Cartesian
basis, given in eq. (\ref{eq:corresp}), the action  of the operators
$a_k^\dagger$ on the present basis can be
calculated for $k=1,\ldots,N-1$:
$$
a^\dagger_k
\left\vert
\begin{array}{c}
\Sigma, \left[n\right]\\
\left[ q \right]
\end{array}
\right>
=
\sqrt{ n_k + q_k/m_k }
\left\vert
\begin{array}{c}
\Sigma', \left[n'\right]\\
\left[ q' \right]
\end{array}
\right>,
$$
where
$$
\begin{array}{c}
n'_\ell=n_\ell \quad q'_\ell = q_\ell \quad  \mbox{for } \ell \ne k, \\
n'_k = n_k + \left[ q_k/m_k\right], \\
\Sigma' = \Sigma + \left[ q_k/m_k\right], \\
q'_k= \mbox{mod }\left(q_k, m_k \right) +1,
\end{array}
$$
while for the operator $a^\dagger_N$ one has
$$
a^\dagger_N
\left\vert
\begin{array}{c}
\Sigma, \left[n\right]\\
\left[ q \right]
\end{array}
\right>
=
\sqrt{ \Sigma - \sum\limits_{k=1}^{N-1}n_k\, + q_N/m_N }
\left\vert
\begin{array}{c}
\Sigma', \left[n'\right]\\
\left[ q' \right]
\end{array}
\right>,
$$
where
$$
\begin{array}{c}
n'_k=n_k,  \qquad q'_k = q_k,  \qquad  \mbox{for }  k=1,\ldots,N-1, \\
\Sigma' = \Sigma + \left[ q_N/m_N\right], \\
q'_N= \mbox{mod }\left(q_N, m_N \right) +1.
\end{array}
$$
Similarly for the operators $a_k$ one can find for $k=1,$ \dots, $N-1$:
$$
a_k
\left\vert
\begin{array}{c}
\Sigma, \left[n\right]\\
\left[ q \right]
\end{array}
\right>
=
\sqrt{ n_k + (q_k-1)/m_k }
\left\vert
\begin{array}{c}
\Sigma', \left[n'\right]\\
\left[ q' \right]
\end{array}
\right>,
$$
where
$$
\begin{array}{c}
n'_\ell=n_\ell,  \qquad q'_\ell = q_\ell,  \qquad  \mbox{for } \ell \ne k, \\
n'_k = n_k + \left[ (q_k-2)/m_k\right], \\
\Sigma' = \Sigma + \left[ (q_k-2)/m_k\right], \\
q'_k= \mbox{mod }\left(q_k-2, m_k \right) +1,
\end{array}
$$
while for the operator $a_N$ one has
$$
a_N
\left\vert
\begin{array}{c}
\Sigma, \left[n\right]\\
\left[ q \right]
\end{array}
\right>
=
\sqrt{ \Sigma - \sum\limits_{k=1}^{N-1}n_k\, + (q_N-1)/m_N }
\left\vert
\begin{array}{c}
\Sigma', \left[n'\right]\\
\left[ q' \right]
\end{array}
\right>,
$$
where
$$
\begin{array}{c}
n'_k=n_k,  \qquad q'_k = q_k,  \qquad  \mbox{for }  k=1,\ldots,N-1, \\
\Sigma' = \Sigma + \left[ (q_N-2)/m_N\right], \\
q'_N= \mbox{mod }\left(q_N-2, m_N \right) +1.
\end{array}
$$

An important difference between the operators $a_k$, $a_k^\dagger$ used
here and the operators $A^{(s)}_i$, ${A^{(s)}_i}^\dagger$ used in
refs \cite{DZ,Ven,MV,ND} has to be pointed out. The operators used
here satisfy the relations
$$ [ a_k, a_l^\dagger ]= {1 \over m_k } \delta_ {kl},$$
i.e. they represent oscillators completely decoupled from each other,
while the operators of refs \cite{DZ,Ven,MV,ND} satisfy the relations
$$ [A_i^{(s)}, {A_j^{(s)}}^\dagger ] = \delta_{ij},\qquad
   [A_i^{(s)}, {A_j^{(s')}}^\dagger ] \neq \delta_{ij}, $$
i.e. they represent oscillators not completely decoupled. Notice that
$(s)$ of refs \cite{DZ,Ven,MV} ($\{\lambda\}$ of ref. \cite{ND}) is analogous
to the $[q]$ used in the present work.

{\bf 3. The 3-dimensional oscillator and relation to the  Nilsson model}

In this section the 3-dim case will be studied in more detail, because of its
relevance for the description of superdeformed nuclei and of nuclear and
atomic clusters. The 3-dim anisotropic oscillator is the basic ingredient of
the Nilsson model \cite{Nil}, which in addition contains a term proportional
to the square of the angular momentum operator, as well as a spin-orbit
coupling term, the relevant Hamiltonian being
$$ H_{Nilsson}= H_{osc} - 2 k \vec{L} \cdot \vec{S} - k \nu \vec{L}^2,$$
where $k$, $\nu$ are constants.
 The spin-orbit term is not needed in the case of atomic
clusters, while in the case of nuclei it can be effectively removed through
a unitary transformation, both in the case of the spherical Nilsson model
\cite{CMQ1,CMQ2,DBM} and of the axially symmetric one \cite{MDM,CVHH}.
An alternative way to effectively remove the spin-orbit term in the
spherical Nilsson model is the q-deformation of the relevant algebra
\cite{DSM}. It should also be noticed that the spherical Nilsson
Hamiltonian is known to possess an osp(1$\vert$2) supersymmetry \cite{BCM}.

In the case of the 3-dim oscillator the relevant operators of eq. (2)  form a
nonlinear generalization of the algebra u(3), the q-deformed version of
which can be found in \cite{STK,SK,L267}.

As we have already seen, to each $\Sigma$ value correspond $m_1 m_2 m_3$
energy eigenvalues, each eigenvalue having degeneracy $(\Sigma+1)(\Sigma+2)/2$.
In order to distinguish the degenerate eigenvalues, we are going to introduce
some generalized angular momentum operators, $L_i$ ($i=1,2,3$),  defined by:
\begin{equation}
L_k =
i \epsilon_{ijk} \left(
\left( a_i \right)^{m_i} \left( a_j^\dagger \right)^{m_j}-
\left( a_i^\dagger \right)^{m_i} \left( a_j \right)^{m_j}
\right).
\label{eq:defang}
\end{equation}
One  can prove that
$$
L_1=i \left( {\cal A}_2 - {\cal A}_2^\dagger \right),
\quad
L_2=i \left( {\cal A}^\dagger_1 - {\cal A}_1 \right).
$$
The following commutation relations can be verified:
\begin{equation}
\left[ L_i, L_j \right] = i \epsilon_{ijk}
\left( F( m_k, U_k+1) - F( m_k, U_k ) \right)L_k.
\label{eq:angmom}
\end{equation}
It is worth noticing that in the case of $m_1=m_2=m_3=1$ one has
$F(1,x)=x-1/2$, so that the above equation gives the usual angular
momentum  commutation relations.

The operators defined in  eq. (\ref{eq:defang}) commute with the
oscillator  hamiltonian
$H$ and therefore conserve the number $\Sigma$ which characterizes the
dimension of the representation. Also these operators do not change the
numbers $q_1$, $q_2$, $q_3$ as we can see from eqs
(\ref{eq:Ak1+}-\ref{eq:Ak1}).
The eigenvalues of these operators can be calculated using Hermite
function techniques.

Let us consider in particular the generalized angular momentum projection:
\begin{equation}
L_3= i
\left( \left(a_1\right)^{m_1} \left(a_2^\dagger\right)^{m_2}-
\left(a_1^\dagger\right)^{m_1} \left(a_2\right)^{m_2} \right).
 \label{eq:l3}
\end{equation}
This acts on the basis as follows
\begin{equation}
\begin{array}{l}
L_3
\left\vert
\begin{array}{c}
\Sigma, \left[n\right]\\
\left[ q \right]
\end{array}
\right>
=\\
=i
\Big(
\sqrt{ F\left(m_1, n_1 + \frac{2 q_{1} -1}{2m_1} \right)
F\left(m_2, n_2 + \frac{2 q_{2} -1}{2m_2}+1 \right)}
\left\vert
\begin{array}{c}
\Sigma, n_{1}-1, n_{2}+1\\
\left[ q \right]
\end{array}
\right>
-\\
-
\sqrt{ F\left(m_1, n_1 + \frac{2 q_{1} -1}{2m_1} +1 \right)
F\left(m_2, n_2 + \frac{2 q_{2} -1}{2m_2} \right)}
\left\vert
\begin{array}{c}
\Sigma, n_{1}+1, n_{2}-1\\
\left[ q \right]
\end{array}
\right>\Big).
\end{array}
\label{eq:L3a}
\end{equation}
This operator conserves the quantum number
$$
j= \frac{n_1+n_2}{2}, \quad j=0,{1\over 2},1,{3\over 2},\ldots
$$
In addition one can  introduce the quantum number
$$
m ={n_1-n_2\over 2}.
$$
One can then replace the quantum numbers $n_1$, $n_2$ by the quantum
numbers $j$, $\mu$, where $\mu$ is the eigenvalue
of the $L_3$ operator. The new representation basis one can label as
\begin{equation}
L_3
\left\vert
\begin{array}{c}
\Sigma,\\
 j, \mu \\
\left[ q \right]
\end{array}
\right>
=
\mu
\left\vert
\begin{array}{c}
\Sigma,\\
 j, \mu \\
\left[ q \right]
\end{array}
\right>.
\label{eq:eigenL3}
\end{equation}
This basis is connected to the basis of the previous section as follows
\begin{equation}
\left\vert
\begin{array}{c}
\Sigma,\\
 j, \mu \\
\left[ q \right]
\end{array}
\right>
=
\sum\limits_{m=-j}^{j}
\frac{ c[j,m,\mu]}
{\sqrt{ [j+m]_1! [j-m]_2! }}
\left\vert
\begin{array}{c}
\Sigma, j+m, j-m\\
\left[ q \right]
\end{array}
\right>,
\label{eq:L3basis}
\end{equation}
where
$$
[0]_k!=1,\quad
[n]_k! = [n]_k [n-1]_k!,
\quad
[n]_k=F\left( m_k, n + \frac{2q_k-1}{2m_k} \right),
$$
and the coefficients  $c[j,m,\mu]$  in eq.
(\ref{eq:L3basis}) satisfy the recurrence relation:
\begin{equation}
\mu c[j,m,\mu] =i
\left(
 [j-m]_2 c[j,m+1,\mu] -[j+m]_1c[j,m-1,\mu]
\right).
\label{eq:L3rec}
\end{equation}
These relations can be satisfied only for special values of
the parameter $\mu$, corresponding to the eigenvalues of $L_3$.
It is worth noticing that in the case
 of $m_1=m_2$, which corresponds to axially symmetric oscillators,
the possible values turn out to be $\mu=-2j, -2(j-1), \ldots, 2(j-1),2j$.
In nuclear physics the quantum numbers $n_{\bot}=n_1+n_2$ and
$\Lambda = \pm n_{\bot}$, $\pm(n_{\bot}-2)$, \dots, $\pm 1$ or 0 are used
\cite{Bohr}. From the above definitions it is clear that $j= n_{\bot}/2$ and
$\mu = \Lambda$. Therefore in the case of $m_1=m_2$, which includes
axially symmetric prolate nuclei with $m_1:m_2:m_3= 1:1:m$, as well as
axially symmetric oblate nuclei with $m_1:m_2:m_3= m:m:1$, the
correspondence between the present scheme and the Nilsson model is clear.

In the case of axially symmetric prolate nuclei ($m_1=m_2=1$) one can easily
check that the expression
$$ L^2 = L_1^2 + L_2^2 + ( F(m_3,U_3+1)-F(m_3,U_3) ) L_3^2 $$
satisfies the commutation relations
$$ [L^2, L_i] =0, \qquad i=1,2,3.$$

{\bf 4. Discussion}

The symmetry algebra of the N-dim anisotropic quantum harmonic oscillator
with rational ratios of frequencies has been constructed by a method
\cite{BDK} of general applicability in constructing finite-dimensional
representations of quantum superinegrable systems.  The case of the
3-dim oscillator has been considered in detail, because of its relevance
to the single particle level spectrum of superdeformed and
hyperdeformed nuclei \cite{Mot,Rae}, to the underlying geometrical
structure of the Bloch-Brink $\alpha$-cluster model \cite{RZ,ZRM}, and possibly
to the shell structure of atomic clusters at large deformations
\cite{Mar,Bul}.  The symmetry algebra in this case is a nonlinear
generalization of the u(3) algebra. For labeling the degenerate states,
generalized angular momentum operators are introduced, clarifying the
connection of the present approach to the Nilsson model.

In the case of the 2-dim oscillator with ratio of frequencies $2:1$
($m_1=1$, $m_2 =2$) it has been shown \cite{BDKL} that the relevant nonlinear
generalized u(2) algebra can be identified as the finite W algebra
W$_3^{(2)}$ \cite{Tj1,Tj2}. In the case of the 3-dim axially symmetric
oblate oscillator with frequency ratio $1:2$ (which corresponds to the case
$m_1 = m_2 =2$, $m_3=1$) the relevant symmetry is related to O(4)
\cite{ND,RSP}. The search for further symmetries, related to specific
frequency ratios, hidden in the general nonlinear algebraic framework given
in this work is an interesting problem.

One of the authors (DB) has been supported by the EU under contract
ERBCHBGCT930467. This project has also been partially supported by the Greek
Secretariat of Research and Technology under contract PENED 340/91.

\vfill\eject

\end{document}